\documentstyle[pra,aps,amsfonts,]{revtex}

\begin{document}

\draft
\twocolumn[\hsize\textwidth\columnwidth\hsize\csname 
@twocolumnfalse\endcsname

\title{Experimental observation of two-dimensional fluctuation magnetization in the vicinity of $T_c$ for low values of the magnetic field in deoxygenated $YBa_2Cu_3O_{7-x}$.}
\author{S. Salem-Sugui Jr.$^{1\ast}$, A. D. Alvarenga$^2$, V. N. Vieira$^3$and O. F. Schilling$^4$}
\address{$^1$ Instituto de Fisica, Universidade Federal do Rio de Janeiro, 21945-970 Rio de Janeiro, Brasil\\
$^2$ Centro Brasileiro de Pesquisas Fisicas, Rua Dr. Xavier Sigaud 150, Rio de Janeiro, Brazil.\\
$^3$ Centro de Ciencias Exatas e Tecnologia, Universidade de Caxias do Sul, RS, Brazil.\\
$^4$ Departamento de Fisica, Universidade Federal de Santa Catarina, Florianópolis, Brazil}
\date{\today}
\maketitle
\begin{abstract}
We have measured isofield magnetization curves as a function of temperature in two single crystal of deoxygenated YBaCuO with $T_c$ = 52 and 41.5 K. Isofield magnetization curves were obtained for fields running from 0.05 to 4 kOe. The reversible region of the magnetization curves was analyzed in terms of a scaling proposed by Prange, but searching for the best exponent $\upsilon$. The scaling analysis carried out for each data sample set with $\upsilon$=0.669, which corresponds to the 3D-xy exponent, did not produced a collapsing of curves when applied to $MvsT$ curves data obtained for the lowest fields. The resulting analysis for the Y123 crystal with $T_c$ = 41.5 K, shows that lower field curves  collapse over the entire reversible region following the Prange's scaling with $\upsilon$=1, suggesting a two-dimensional behavior. It is shown that the same data obeying the Prange's scaling with $\upsilon$=1 for crystal with $T_c$ = 41.5 K, as well low field data for crystal with $T_c$ = 52 K, obey the known two-dimensional scaling law obtained in  the lowest-Landau-level approximation. 
\end{abstract}
]
We report on isofield magnetic measurements performed on two single crystals of deoxygenated YBaCuO with $T_{c}$ = 52 and 41.5 K. The experiment address effects of diamagnetic fluctuations occurring near the upper critical temperature, $T_{c2}(H)$ in deoxygenated YBaCuO for low values of the magnetic field.\\
Superconducting diamagnetic fluctuations above and below $T_c$ have been the issue of extensive investigations since the discovery of the high-$T_c$ materials. These materials due to their high value of $T_c$, low coherence lengths and layered structure (anisotropy), display a much larger fluctuations effects than the low $T_c$ superconductors \cite{lee,welp}. For high-$T_c$ materials the $\delta T$ region around $T_c$ where critical fluctuations occur can be of the order of 1 K \cite{lee}.\\
From the theoretical point of view diamagnetic fluctuations in layered materials were previously treated in detail by Klemm et al., \cite{klemm} and by Gerhardts \cite{ger}. These Authors \cite{klemm} calculated the fluctuation magnetization as a function of magnetic field and temperature and obtained scaled plots predicting a field-induced dimensional crossover and the behavior of magnetization curves for three-dimensional as well two-dimensional systems. \\
For YBaCuO, many experiments give evidence for the existence of 3D-XY critical fluctuations in relatively large magnetic fields \cite{inderhees,salamon,hubbard,kamal,yeh,moloni}. On the other hand, similar measurements shown to obey also a scaling law predicted by lowest-Landau-level (LLL) fluctuations theory \cite{welp}. From the theoretical point of view, the fact that similar sets of data obey both scaling laws is conflicting \cite{lawrie}. For deoxygenated YBaCuO the removal of oxygen increases the anisotropy. In this case, a reduction of the size of the $\delta T$ region where critical flutuations occur is expected\cite{hubbard}. On the other hand, experiments \cite{moloni} performed in deoxygenated YBaCuO films have shown that 3D-XY scaling is obeyed in a field region of 10-50 kOe, whereas LLL scaling is obeyed at higher fields. It appears that there is a shortage of experimental data in the literature displaying low-field 3D-XY scaling particularly for fields in which $T_{c2}(H)$ is close to $T_{c}$. It is thought that the superconducting transition for zero field occurring at $T_{c}$ belongs to the same class of 3D-XY criticality as superfluid liquid He \cite{fisher}. Therefore, one may expect that fluctuations occurring for low fields where $T_{c2}(H)$ is close to $T_{c}$ should obey 3D-XY critical scaling. \\
We address the above issue, by obtaining magnetization versus temperature curves for deoxygenated YBaCuO samples with magnetic fields applied parallel to the c axis, running from 0.05 kOe to 4 kOe . Analysis of the data  according to the 3D-XY critical scaling shows that the lowest field curves do not obey this scaling indicating that the fluctuation magnetization for these samples do not belong to the 3D-XY class. The resulting analysis for the studied Y123 crystals show that the fluctuation magnetization near $T_c$ for low fields have a two dimensional character as low field curves collapse following the two-dimensional lowest-Landau-level scaling. \\
The measured samples were: two single crystals of $YBa_{2}Cu_{3}O_{7-x}$
with critical temperatures, $T_c$= 52 K (x=0.5), and 41.5 K (x=0.6). The single crystals of YBaCuO were grow at Argonne National Laboratory \cite{veal}, and have approximate dimensions: 1x1x0.2mm (sample with $T_c$=52 K, m=1 mg)) and 1x1x0.1mm (sample with $T_c$=41.5 K, m=0.5 mg) with the $c$ axis along the shorter direction, and each exhibited sharp, fully developed transitions ($\Delta T_{c}\simeq1K$).\\
A commercial magnetometer based on a
superconducting quantum interference device (SQUID) (MPMS Quantum Design)
was utilized. The scan length was 3 cm, which minimized field inhomogeneities. Experiments were conducted by
obtaining isofield magnetization (M) data as a function of temperature,
producing $M-vs-T$ curves, with values of field running from 0.05 kOe to 8 kOe. Magnetization data were always taken after cooling the sample below $T_{c}$,
in zero magnetic field, zfc. In fact, all measurements were obtained in the presence of the Earth magnetic field. After cooling to the desired temperature, a magnetic field was carefully applied (without
overshooting), always along the $c$ axis direction of the samples, and $M-vs-T$
curves were obtained by heating the sample up to temperatures well above $T_{c}$, for fixed $\Delta T$ increments. We also obtained field-cooled curves, corresponding to cooling the sample from above to below $T_{c}$, in an applied magnetic
field. This procedure allowed determining the reversible (equilibrium)
magnetization. A careful background correction was performed for each $MvsT$ curve. For the Y123 sample with $T_c$=52 K, this correction was of the type $M(T)=A(H)$ where $A(H)$ for each curve is a constant diamagnetic value reached just above $T_c$ extending over the entire measured region above $T_c$ (10K). The value of $A(H)$ for the later satisfy $A(H)=6.16 10^{-6}-4.8 10^{-8}H$ with H in Oe. The same trend was observed for the Y123 sample with $T_c$=41.5 K, where the correction $M(T) = A(H)$, with $A(H)=1.8 10^{-7}-1.4 10^{-8}H$ but only up to $H=1 kOe$. Above $H=1kOe$ an additional term $C(H)/T$ had to be considered with $C(H)$ very small.\\
To eliminate the possibility of studying magnetization curves which show high-field diamagnetic fluctuations, we limited our analysis on data obtained for fields up to 4 kOe. The existence of high-field diamagnetic fluctuations in high-Tc superconductors is in most of the cases indicated by a field independent magnetization value occurring close to $T_c$, which appears as a crossing of different $MvsT$ curves \cite{tesa,said1,said2}. By plotting together our $M-vs-T$ curves for the Y123 samples such crossing point appears for fields greater than 1 kOe.\\
The initial scaling analysis performed on the data is based on the exact expression obtained by Prange \cite{prange} for the fluctuation magnetization:\\
$M=-2\pi^{1/2} kT\phi_{0}^{-3/2}H^{1/2}g(x)$
where $k$ is the Boltzman constant, $\phi_{0}$ is the quantum flux and g(x) is a universal function of the variable 
$x=(\frac{dH_{c2}}{dT})_{T_c}\frac{T-T_c}{H^{1/2\upsilon}}$ 
where $\upsilon$=1/2 for Gaussian fluctuations. It is worth mentioning that this scaling law has been shown to apply for many conventional superconductors in the presence of weak magnetic fields and for temperatures close to $T_c$ \cite{gollub}. \\
The initial motivation was to verify if data obtained for the lowest fields obey (or not) the 3D-XY scaling. As in Ref. , the scaling analysis consisted on plotting values of $M/H^{0.5}$ against values of $(T/T_{c}-1)/H^{0.747}$ where $T_c=T_{cxy}$ is a fitting parameter, chosen to produce the best collapsing of the curves. The exponent 0.747=$1/2\upsilon$ where $\upsilon$=0.669 is the 3D-XY critical exponent \cite{leguillou}. Figures 1a and 1b show the results of the scaling analysis performed on the reversible data, which is presented in the inset of each figure. For better visualization of the scaling and of the curves obtained for low fields Figs. 1a and 1b include some zero-field-cooled data in the irreversible region also. The results shown in Fig.1 where obtained with values of $T_{cxy}$ (the fitting parameter) close to the respective measured $T_c$. We also allow small variations (of the order of $5\%$)in the value of the exponent (0.747), which did not produce better results than those presented.\\ An eye inspection of Fig. 1a (Y123 with $T_{cxy}$=51.8 K) and Fig. 1b (Y123 with $T_{cxy}$=41.6 K), reveals that the lowest field curves fail to collapse onto higher field curves. This fact was observed in Fig. 1a, where the curves with 0.05 and 0.1 kOe are separated from the collapsing curve formed by the curves with higher fields, $H\geq$0.25 kOe. A similar tendency was observed in Fig. 1b, which shows the low field curves with H equal 0.05, 0.1, 0.25 and 0.4 kOe spread apart from the collapsing curve formed by the curves with 0.8, 1, 2, and 4 kOe. Despite the fact that Fig. 1a shows the curves with 0.05 and 0.1 kOe collapsing, it is inappropriate to admit the existence of a detached collapsing curve formed with lower field curves only. It is shown below that low field curves for this sample, including 0.05, 0.1, 0.25 and 0.5 kOe, obey a two-dimensional scaling. It is important to point out that, despite the good collapse of high field curves shown in Fig.1a and 1b, the lowest field curves are those expected to follow the 3D-XY scaling and fail to follow. The latter indicates that diamagnetic fluctuation magnetization induced by field in the studied samples do not belong to the class of 3D-XY.\\
As an attempt to study the fluctuation magnetization in the vicinity of $T_c$ and to search for a possible power law behavior with magnetic field, we have tried an analysis based on the Prange \cite{prange} scaling discussed above. We plotted $MH^{-0.5}/T$ vs $H$ calculated at fixed temperatures. For Y123 with $T_c$=41.5 K, the plots produced power law curves with $H^{-0.55}$calculated at $T=T_c$, and $H^{-0.44}$ for $(T-T_{c})$= -0.2 K. For Y123 with $T_c$=52 K it has only been possible to obtain a power law with $H^{-0.74}$ for $(T-T_{c})$= -0.5 K (T=51.5K), which is rather far from $T_c$=52 K. For the latter case, the scaling analysis fails to produce a power law at $T_c$ (values of magnetization at $T_c$ and close, are extremelly small for this sample, of the order of $10^{-6}$ emu). The power law behavior found is mostly determined by the tendency for divergence that occurs as the field approaches zero. This fact limits the analysis to a region below $T_c$  $\sim0.5 K$ because for temperatures for which $(T-T_{c})\succ -0.5K$ magnetization values for the lowest field curve are in the irreversible region and cannot be analyzed. The exponents found (with the exception for the sample with $T_c$=52 K whose exponent was obtained 0.5 K far away from $T_c$) are then expected to be more appropriate for curves obtained with low fields. We note that the exponents of $H$ found for the sample with $T_c$ = 41.5 K are rather different than -0.74, and this led us to repeat the analysis performed in Fig. 1b starting with the exponent $1/2\upsilon$=0.55. Despite the scaling analysis of data in Fig. 1b for $1/2\upsilon$=0.55 having produced a collapse of the low field curves, the best collapse was obtained with $1/2\upsilon$=0.5, which corresponds to the exponent $\upsilon$=1, and a fitting paramenter $T_{c}$=41.75 K (slightly larger than the experimental value). Before presenting the latter scaling analysis, it should be noted that the plot with $1/2\upsilon$=0.5, $M/H^{0.5}$ vs $T/T_{c}-1)/H^{0.5}$ suggests a 2D behavior since it is very similar to the plot $M/(TH)^{0.5}$ vs $(T-T_{c}(H))/(TH^{0.5})$ which corresponds to the two-dimensional scaling law obtained in the lowest-Landau-level approximation \cite{welp}, 2D-LLL. It is worth mentioning that despite the 2D-LLL scaling law was obtained in the lowest-Landau-level approximation, the fact that $MvsT$ curves obey this scaling indicates that the fluctuation magnetization has a two-dimensional nature independently of the value of the applied magnetic field. Figure 2 shows the results of the 2D-LLL scaling analysis performed on the low field data for sample with $T_c$ =41.5 K where all four low field curves (H=0.05, 0.1, 0.25, 0.4 kOe), which did not collapse in Fig. 1b, appear now fully collapsed (extending over the entire reversible region). In the 2D-LLL analysis $T_{c}(H)$ is the fitting parameter which is adjusted for each $MvsT$ curve to produce the best collapsing curve shown in Fig. 2. The scaling analysis produced consistent values of $T_{c}(H)$=41.745 (H=0.05 kOe), 41.74 (H=0.1 kOe), 41.73 (H=0.25 kOe) and 41.71 (H=0.4 kOe). We note that higher field curves also follow the scaling of Fig. 2 (not shown), collapsing with the low field curves, but only within a narrow temperature region close to $T_{c}(H)$. Since the collapsing of the curves shown in Fig. 2 occur within a 2D critical region, one may infer a $\delta T$ critical region below $T_c$ of the order of 1K for this sample. This temperature region includes the full reversible region of $MvsT$ curves obtained with low fields ($H\leq 0.4 kOe$), and it is likely that the irreversibility line in this field region should exhibit 2D behavior. Unfortunately it is not possible to check that with only 4 points in a $H-vs-T$ diagram. The inset of Fig. 2 shows values of the normalized magnetization, $(M(T_c)shc/2ek_BT_c)$, extracted at $T_c$, plotted as a function of H as in Fig. 5 of Ref.3 , where $s=8.2 \AA$ is the distance between layers, $hc/2e=\phi_0$ the quantum flux and $K_B$ is the Boltzmann constant. Values of $M(T_c)$ in this inset produced the power law curve with $H^{-0.55}$ after performing the $MH^{-0.5}/T$ vs $H$ plot discussed above. The purpose of the plot in the inset of Fig. 2 was to compare its behavior with the plot presented in Fig. 5 of Ref.3 which was drawn for two dimensional magnetization thin films. In our curve values of H are much smaller than values of Fig. 5 in Ref. 3 and we only have data up to 8 kOe, but from the tendency of the curve for high fields, it is likely that the magnetization at T=$T_c$ should drop as the field increases following the expected 2D behavior. We note that the curve in the inset of Fig. 2 do not approach the Prange value as H goes to zero. We do not have an explanation for that. We mention that the $MvsH$ curve obtained at $T-T_c$=-0.2 K is very similar to the curve in the inset of Fig. 2. The good results of the 2D analysis of Fig. 2 motivated us to extend the 2D-LLL analysis to the data of the crystal with $T_c$= 52 K also. Figure 3 shows the results of the two-dimensional-LLL analysis performed on the low field data for the sample with $T_c$=52 K, where low field curves with H=0.05, 0.1, 0.25 and 0.5 kOe appear  fully collapsed within a $\delta T \approx 0.7 K$ region below $T_{c}(H)$. The reversible data corresponding to the $MvsT$ curves with H=0.05 and 0.1 kOe appear fully collapsed in Fig. 3. It should be noted that in Fig.1b (3D-XY analysis) the curves with H=0.25 and 0.5 kOe are totally isolated from the curves with H=0.05 and 0.1 kOe. As in Fig. 2, the scaling analysis of Fig. 3 produced consistent values of $T_{c}(H)$=52.2 (H=0.05 kOe), 52.195 (H=0.1 kOe), 52.175 (H=0.25 kOe) and 52.15 (H=0.5 kOe). It should be mentioned that the obtained values of $T_c(H)$ for both samples produced values of $|dH_{c2}/dT|$ of the order of 10 kOe/K, which are in good agreement with values obtained for deoxygenated crystals with same content of oxygen \cite{JLTP}.\\
Finaly, we estimate the value of the parameter \cite{ger} $r=8(m/M)[\varsigma _{GL}(0)/(\pi s)]^2$ for our samples, where m and M are the effective mass of the quasi-particles along the layers plane and perpendicular to the layers respectively, $\varsigma _{GL}(0)^2= \phi_{0}/(2\pi T_c |dH_{c2}/dT|$ and $s=8.2\AA$. The parameter $r$, first defined in Ref.3, carries important information about the dimensionality of the system. It is worth mentioning that the above expression for $r$ defined in Ref.4  coincides with the expression for $r$ defined by Klemm et al.\cite{klemm} for systems in the dirty limit when calculated at $T=T_c$. The calculated values are: $r$=0.00086 for the crystal with $T_c$=41.5 K where we used $|dH_{c2}/dT|=10 kOe/K$ \cite{JLTP}, $\sqrt{M/m}=105$ \cite{chien} and $\varsigma _{GL}(0)=28\AA$, and $r$=0.0027 for the crystal with $T_c$=52K where we used $|dH_{c2}/dT|=7 kOe/K$ \cite{JLTP} and $\sqrt{M/m}=65$ \cite{chien} and $\varsigma _{GL}(0)=31\AA$. The listed $|dH_{c2}/dT$ values extracted from Ref.21 were obtained for the same crystal of this work with $T_c$=52K and a YBaCuO crystal with approximately same value of $T_c$ ($T_c$=41.5K) belonging to the same batch of the sample used in this work (samples grown by Dr. B. Veal), and values of $\sqrt{M/m}$ extracted from Ref.22 were obtained for samples with approximately same content of oxygen grown at the same laboratory \cite{veal}. It is important to mention that the estimated values of $r$ are below the values appearing in the calculated plot of $H/H_s$ versus $r$ ($H_s=H_{c2}(0)$) of Fig. 7 in Ref.3. This important plot shows the behavior of the field-induced dimensional crossover line as a function of $r$ and it is worth mentioning that the Authors of Ref.3 have suggested that systems with very low values of $r$, may enter the superconducting transition displaying 2D fluctuation for low fields. The latter observation indicates that the two-dimensional fluctuation magnetization behavior observed near $T_c$ for our samples is in agreement with the estimated low values of $r$.\\

We thank L. Moriconi for helpfull discussions. We also thank B. Veal from Argonne National Laboratory, for kindly providing the samples. This work was partially supported by CNPq, Brazilian Agency.\\
$\ast$ Corresponding author. E-mail: said@if.ufrj.br

\begin{figure}[htb]
\vspace*{4cm}
\special{eps: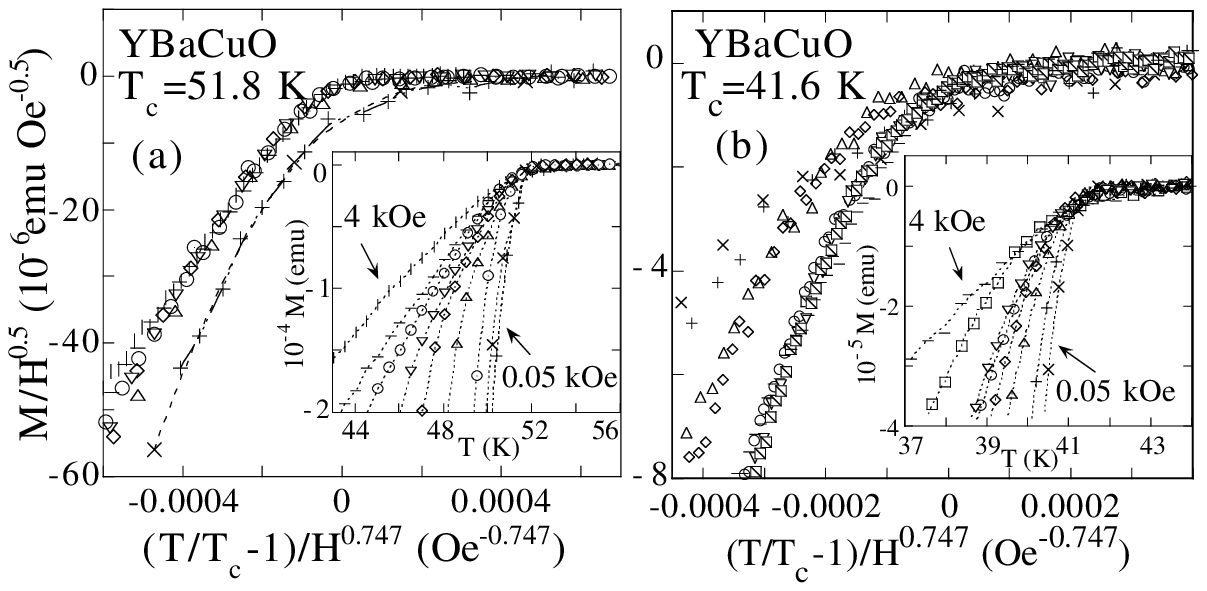 x=8.5cm y=4,5cm}
\caption{$M/H^{0.5}$ versus $(T/T_{c}-1)/H^{0.747}$ for: a) Y123 with $T_c$=52 K. The inset shows MvsT curves used in the main figure, obtained for H = 0.05, 0.1, 0.25, 0.5, 0.75, 1, 1.5, 2, 4 kOe. b) Y123 with $T_c$=41.5 K. The inset shows MvsT curves used in the main figure, obtained for H = 0.05, 0.1, 0.25, 0.4, 0.8, 1, 2, 4 kOe.} 
\end{figure}

\begin{figure}[htb]
\vspace*{8cm}
\special{eps: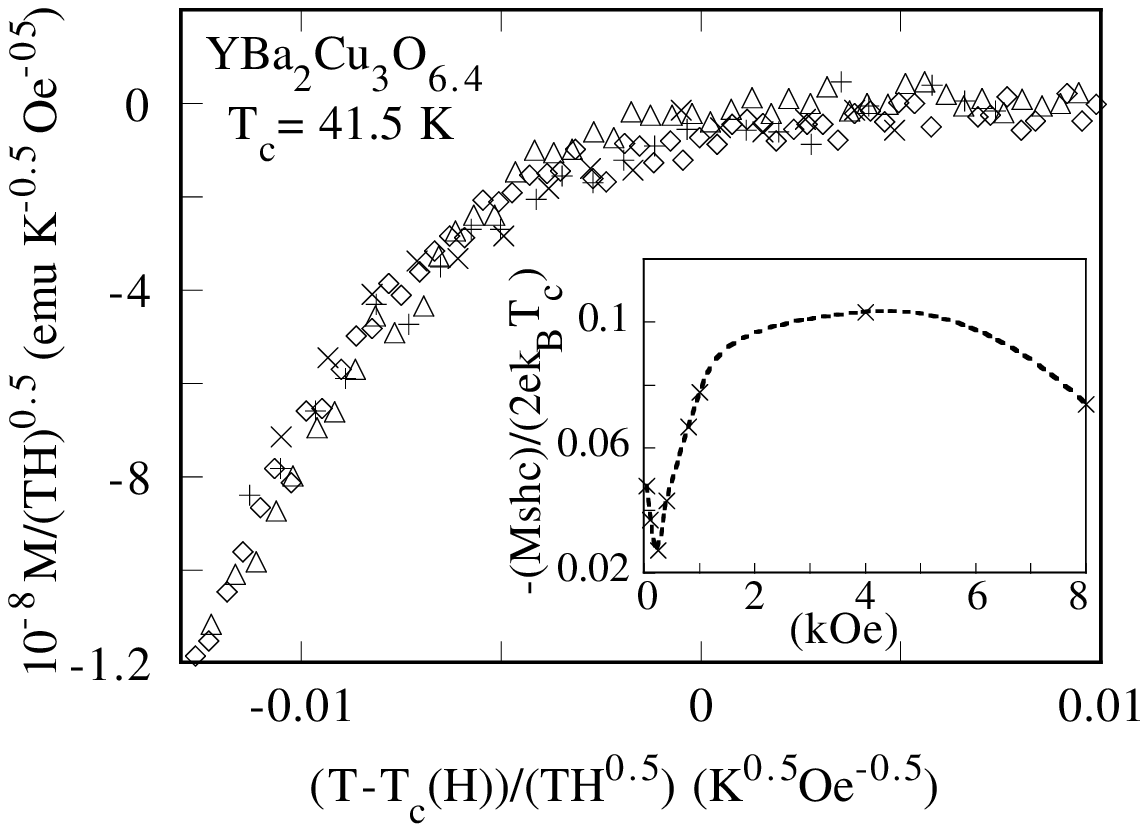 x=8.5cm y=7cm}
\caption{$M/(TH)^{0.5}$ versus $(T-T_{c}(H))/(TH)^{0.5}$ for crystal with $T_c$=41.5 K, obtained from MvsT curves with H=0.05, 0.1, 0.25, 0.4 kOe, where $T_{c}(H)$ is the fitting parameter. The inset shows $(M(T_c)shc/2ek_BT_c)$ plotted against H at $T=T_c$.} where $M(T_c)$ is in emu units.
\end{figure}

\begin{figure}[htb]
\vspace*{8cm}
\special{eps: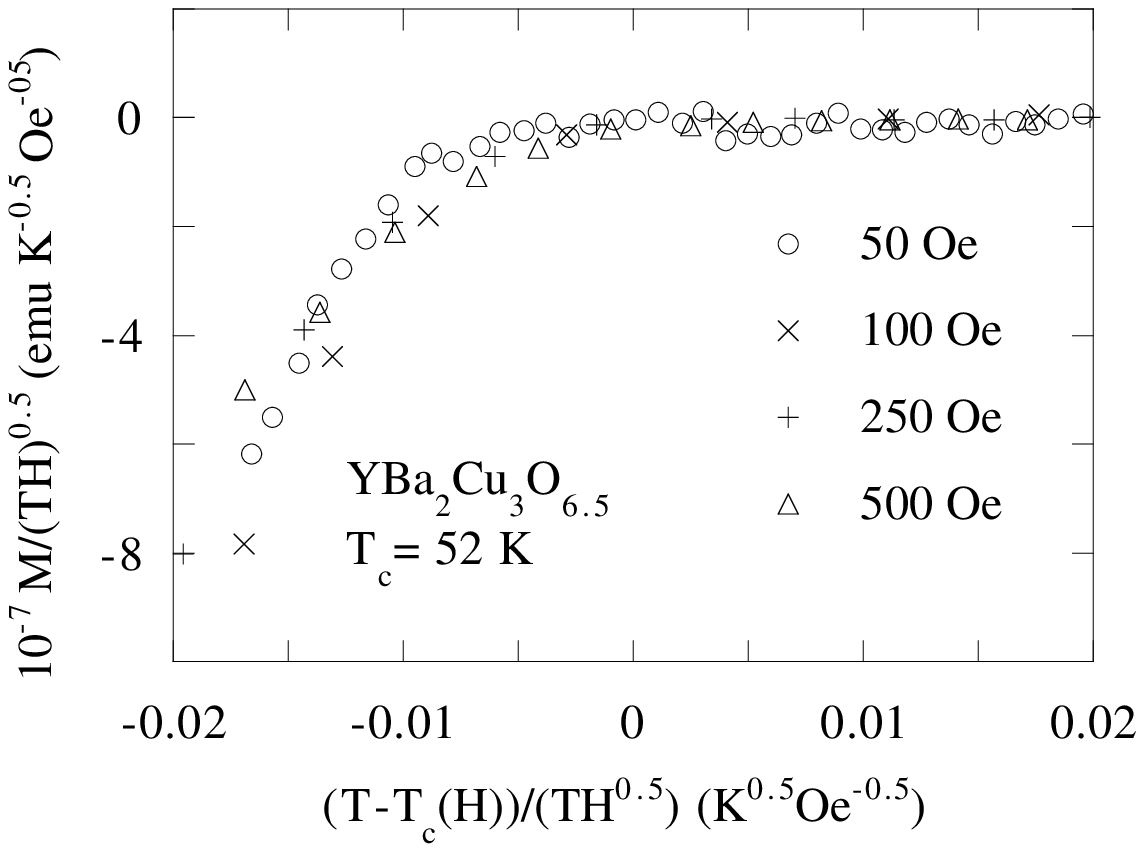 x=8.5cm y=7cm}
\caption{$M/(TH)^{0.5}$ versus $(T-T_{c}(H))/(TH)^{0.5}$ for crystal with $T_c$=52 K, obtained from MvsT curves with H=0.05, 0.1, 0.25, 0.5 kOe, where $T_{c}(H)$ is the fitting parameter.} 
\end{figure}

\end{document}